\documentclass[aps,prl,twocolumn,showpacs,floatfix,amsmath,amssymb,superscriptaddress,showkeys]{revtex4-2}
\usepackage{graphicx}
\usepackage{epstopdf}
\usepackage{color}
\usepackage{hyperref}
\usepackage{bm}
\usepackage{printlen}
\usepackage{units}
\usepackage{bbm,pifont}
\usepackage{tikz}
\usepackage{amsmath}
\usepackage{amssymb}
\usepackage{amsthm}
\usepackage{commath}
\usepackage{amsfonts}
\usepackage[utf8]{inputenc}
\usepackage{dcolumn}
\usepackage{epsfig}
\usepackage{latexsym}
\usepackage{xcolor}
\usepackage{subfigure}
\usepackage{array}
\usepackage{bbm}
\usepackage{hyperref,hypcap}
\usepackage{cancel}
\usepackage{ulem,dsfont}
\usepackage{braket}

\begin{document}

\title{Adiabatic Pumping of Orbital Magnetization by Spin Precession}

\author{Yafei Ren}
\affiliation{Department of Physics and Astronomy, University of Delaware, Newark, Delaware 19716, USA}
\affiliation{Department of Materials Science and Engineering, University of Washington, Seattle, Washington 98195, USA}

\author{Wenqin Chen}
\affiliation{Department of Physics, University of Washington, Seattle, Washington 98195, USA}

\author{Chong Wang}
\affiliation{Department of Materials Science and Engineering, University of Washington, Seattle, Washington 98195, USA}

\author{Ting Cao}
\affiliation{Department of Materials Science and Engineering, University of Washington, Seattle, Washington 98195, USA}

\author{Di Xiao}
\affiliation{Department of Materials Science and Engineering, University of Washington, Seattle, Washington 98195, USA}
\affiliation{Department of Physics, University of Washington, Seattle, Washington 98195, USA}

\date{\today}

\begin{abstract}
We propose adiabatic pumping of orbital magnetization driven by coherent spin precession, {facilitating the rectification of this precession. 
The orbital magnetization originates from the adiabatic evolution of valence electrons with a topological bulk contribution expressed as a Chern-Simons form. When the precession cone angle of spin $\bm{S}$ is small, the resulting magnetization is proportional to $\bm{S}\times \dot{\bm{S}}$, contributing to the magnon Zeeman effect. With a large cone angle, the magnetization can reach its natural unit, $e/T$, in an antiferromagnetic topological insulator with $e$ as the elementary charge and $T$ as the precession period. This significant magnetization is related to the global properties of the electronic geometric phases in the parameter space spanned by $\bm{S}$ and momentum $\bm{k}$. When the pumped magnetization is inhomogeneous, induced by spin textures or electronic topological phase domains, a dissipationless charge current is also pumped. At last, we discuss the boundary contributions from the spin-driving edge states, which are intricately linked to the gauge-dependent quantum uncertainty of the Chern-Simons form.}
\end{abstract}

\maketitle

Orbital magnetization has been a focus of recent research in condensed matter physics. 
It plays a crucial role in the orbital Chern insulator, a topological phase driven by strong correlation where the Chern number is closely related to the orbital magnetization~\cite{sharpe2019emergent, serlin2020intrinsic, chen2020tunable, liu2019quantum, ren2021orbital}. Electric manipulations of the orbital magnetization and thus topological phase have been reported~\cite{polshyn2020electrical, zhu2020voltage, su2020current}.
An electric field can also induce orbital magnetization in the axion insulator with topological magnetoelectric effect~\cite{qi2008topological, xiao2009polarization, essin2009magnetoelectric, mong2010antiferromagnetic, he2022topological} or in transition metals through orbital Hall effect~\cite{bernevig2005orbitronics,kontani2009giant,go2018intrinsic, choi2023observation}, which revives the emerging field of orbitronics~\cite{bernevig2005orbitronics, go2017toward, phong2019optically, go2021orbitronics,sala2023orbital} that encodes information with the orbital degree of freedom. In addition to the electric means, a spin-based scheme to engineer the orbital magnetization is fundamentally interesting and practically important by bridging the fields of orbitronics and spintronics~\cite{vzutic2004spintronics,baltz2018antiferromagnetic,vsmejkal2018topological,go2023orbital, hayashi2023observation}.

This Letter proposes pumping orbital magnetization in magnetic insulators using coherent spin precession. The orbital magnetization of electrons can have intra-atomic and inter-atomic contributions. While the former is partially or completely quenched by the crystal field~\cite{go2021orbitronics}, the latter however can be large and is closely related to the electronic geometric phases~\cite{go2021orbitronics, xiao2009polarization, essin2009magnetoelectric}. Therefore, we focus on topological systems with prominent geometric phase effects~\cite{hohenadler2011correlation, hohenadler2012quantum, hutchinson2021analytical, liu2024gate, gong2019experimental, otrokov2019prediction, chang2023colloquium, chen2024antiferromagnetic}. Such systems exhibit additional advantages. First, abundant 2D van der Waals layers with nontrivial topology and/or magnetism have been suggested by recent theoretical and experimental progresses~\cite{hohenadler2011correlation, hohenadler2012quantum, hutchinson2021analytical, liu2024gate, gong2019experimental, otrokov2019prediction, chang2023colloquium, chen2024antiferromagnetic, gong2019two, cortie2020two, ningrum2020recent, yao2021recent, olsen2019discovering, marrazzo2019relative, choudhary2020computational}, which allows the realization of our proposals by using the 2D atomic legos~\cite{geim2013van}.
Second, the pumping is universal for both ferro- and anti-ferromagnet. Particularly, the pumping in antiferromagnet topological insulators can reach its topological limit, $e/T$, which could provide a starting point for the study of orbital physics with spin dynamics in topological antiferromagnet~\cite{bernevig2022progress, bonbien2021topological} that shows potential applications in orbitronics and antiferromagnetic spintronics~\cite{baltz2018antiferromagnetic, vsmejkal2018topological}.

In the following, {we explore the magnetization pumping in an antiferromagnetic topological insulator. The precession can be driven by a microwave in magnetic resonance or spin torque induced by a current. A large precession cone angle is achievable with easy-plane anisotropy.
We focus on the Kane-Mele-Hubbard model with spontaneous magnetic order where the dynamics of valence electrons are capured by a mean-field Hamiltonian. We explore how the magnetization varies as the precession cone angle increases and elaborate the effect of geometric phases. We then discuss the pumping of a dissipationless charge current that arises from the inhomogeneous orbital magnetization in the presence of spin textures or topological phase domains. At last, we address the contributions from boundaries and their relationship with that from the bulk.}

{\textit{\textbf{Coherent spin precession.---}} Coherent spin precession is a crucial ingredient in spintronics and magnonics for generating spin and magnon current~\cite{han2023coherent}. Various methods are routinely used to generate the coherent precession, including microwave-induced magnetic resonance or spin-orbit torques induced by current in a neighboring layer, where the spin dynamics are described by the Landau-Lifshitz-Gilbert (LLG) equation~\cite{ellis2015landau}
\begin{align}
    \dot{\bm{S}}_i = \gamma (\bm{H}_i^{\rm eff}+\bm{H}_i^{\rm T})\times \bm{S}_i + \alpha \bm{S}_i \times \dot{\bm{S}}_i
\end{align}
where $\gamma$ is the gyromagnetic ratio, $\bm{S}_i$ represents the spin orientation at the $i$-th site, $\bm{H}_i^{\rm eff}$ stands for the effective magnetic field for $\bm{S}_i$ in the absence of external stimuli, $\bm{H}_i^{\rm T}$ is the external field from microwave or spin-orbit torque, and $\alpha$ is the Gilbert damping. A sustained precession can be achieved when external stimuli offset the energy loss due to damping~\cite{SM}.

The precession angle varies in a broad range for different systems. With easy-axis anisotropy, the equilibrium spin orientation is locked to the easy-axis direction and the cone angle of the precession can reach a few degrees around the easy axis~\cite{kiselev2003microwave, han2023coherent}. With easy-plane magnetic anisotropy, the spin can coherently rotate in the easy plane with a much larger cone angle, approaching 90 degrees, driven by spin-orbit torques~\cite{cheng2015ultrafast, cheng2016terahertz, takeuchi2021chiral, yan2022quantum}. 
}

\begin{figure}
\includegraphics[width=1\columnwidth]{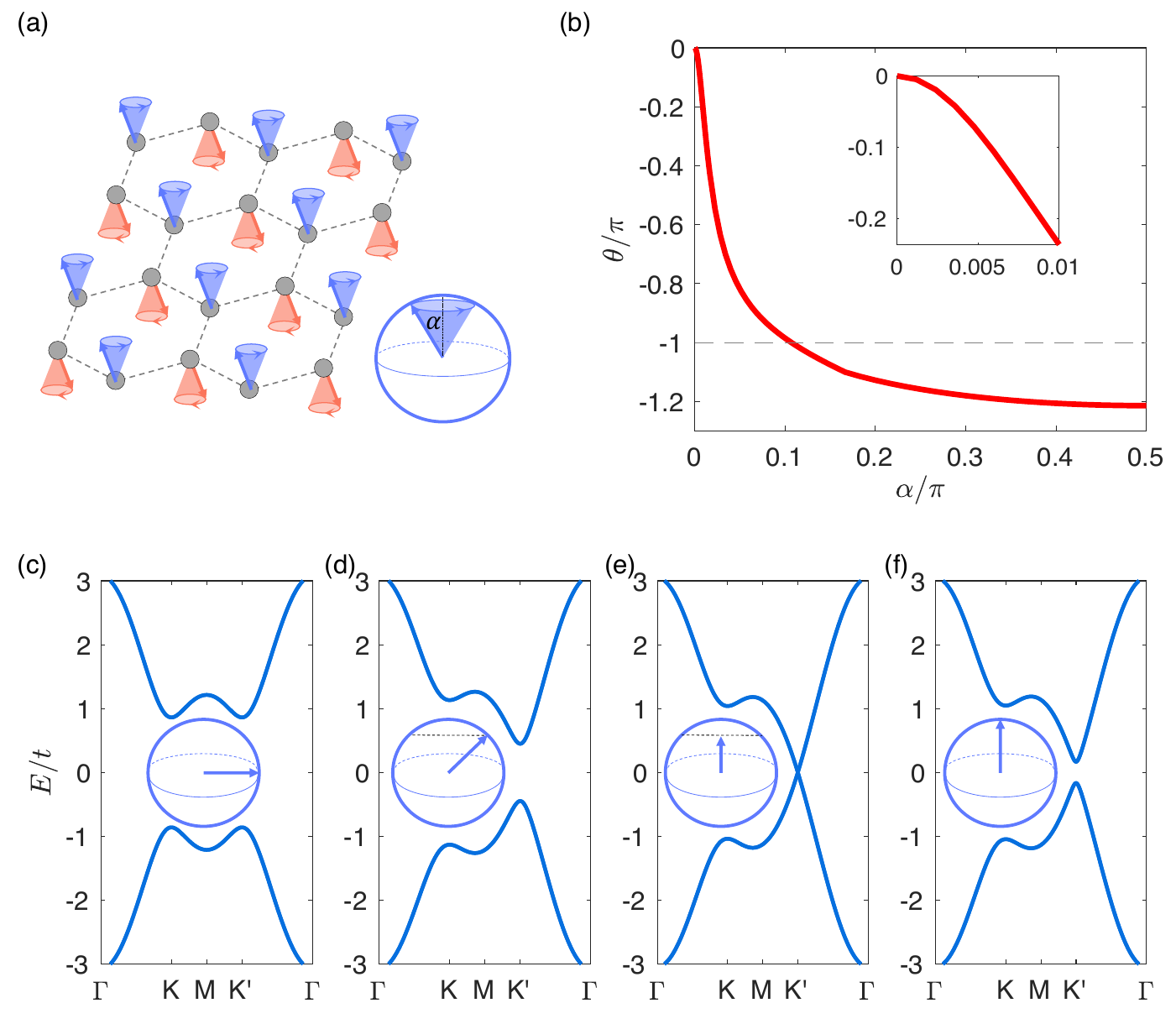}
\caption{(a) Schematic of an easy-axis antiferromagnetic topological insulator with spin precession. (b) Chern-Simons form vs the spin precession angle $\alpha$. Inset: schematic of the $\alpha$. The arrow indicates the orientation of the N\'eel vector $\bm{S}$. (c)-(f) Energy band structure at different spin configurations. Insets: N\'eel vectors.}
\label{Fig4}
\end{figure}

{\textit{\textbf{Model system.---}} We focus on antiferromagnetic topological insulators where spin magnetization vanishes and the topological electronic structure can enhance the orbital-magnetization pumping. A prototype system is the Kane-Mele Hubbard model~\cite{hohenadler2011correlation, hohenadler2012quantum, hutchinson2021analytical, liu2024gate} where N\'eel-type antiferromagnetic order forms spontaneously with easy-plane anisotropy. The electronic system is described by the following mean field Hamiltonian~\cite{hutchinson2021analytical}} 
\begin{align}
H_{\rm AF} = H_{\rm KM} + \lambda \sum_i \bm{S}_i(t) \cdot \bm{s}_i
\end{align}
where $H_{\rm KM}$ represents the Kane-Mele model Hamiltonian for the valence electrons 
\begin{align}
    H_{\rm KM} = \sum_{\bm{k}} f_x \sigma_x s_0 + f_y \sigma_y s_0 + f_z \sigma_z s_z
\end{align}
where $\bm{\sigma}$ and $\bm{s}$ are Pauli matrices of sublattice and spin, separately. Here, $f_x = 1+2\cos q_y \cos q_x$, $f_y = 2\sin q_y \cos q_x$, and $f_z = -2\gamma (\sin 2 q_x - 2 \cos q_y  \sin q_x)$ with $q_y=\frac{3}{2}k_y$ and $q_x=\frac{\sqrt{3}}{2} k_x$.
The lattice constant $a$ and inter-site hopping $t$ are set to be 1, $\gamma$ characterizes the next-nearest-neighbor hopping from the intrinsic spin-orbit coupling that opens a band gap of $6\sqrt{3}\gamma$ at the Dirac points $K/K'$.
The unit vector $\bm{S}_i$ represents the spin orientation at site $i$ formed spontaneously due to the Hubbard interaction~\cite{hutchinson2021analytical} and $\lambda$ indicates the exchange field induced by the magnetic order.
{In the absence of spin precession, the electronic structure and band topology strongly depend on the direction of the N\'eel vector. When the N\'eel vector lies in the basal plane, the band structure, as shown in Fig.~\ref{Fig4}(c), exhibits degenerate valleys at $K$ and $K'$ points. Tilting the N\'eel vector out of the plane lifts this valley degeneracy, as plotted in Fig.~\ref{Fig4}(d). As the in-plane spin component vanishes while keeping the out-of-plane component fixed, the band gap closes, as shown in Fig.~\ref{Fig4}(e). The gap reopens as the out-of-plane spin component increases to the unit as shown in Fig.~\ref{Fig4}(f).
Due to the parity-time ($\mathcal{P}\mathcal{T}$) symmetry, each band is doubly degenerate, ensuring vanishing static spin and orbital magnetization. However, nonzero orbital magnetization appears in the presence of spin precession, represented by the periodically varying $\bm{S}_i(t)$.}

\textit{\textbf{Formalism.---}} 
{As the precession frequency typically ranges from GHz to THz that is smaller than the typical band gaps of antiferromagnetic topological insulators, the valence electrons are thus pumped adiabatically.} 
The orbital magnetization from adiabatic pumping can be obtained by studying the current density $\bm{j}$ induced by time-varying parameters, e.g., the spin configuration $\bm{S}(\bm{r},t)$ that vary slowly in space and time~\cite{trifunovic2019geometric, dong2018geometrodynamics, xiao2021adiabatically, ren2021phonon}. 
By comparing the current density with $\bm{j}=\partial_t\bm{P} + \nabla\times \bm{M}$, the theoretical formulas of orbital magnetization $\bm{M}$ can be extracted~\cite{trifunovic2019geometric, dong2018geometrodynamics, xiao2021adiabatically, ren2021phonon}. The spatial varying of $\bm{S}(\bm{r},t)$ is assumed for the convenience of derivation. After obtaining the formalism of $\bm{M}$, one can calculate the magnetization under $\bm{S}$ that is spatially uniform and temporally periodic. As we focus on 2D systems, $\bm{M}$ only has a nonzero $z$-component.
The orbital magnetization has two contributions and we focus on the topological part that is more relevant to the physics of interest~\cite{trifunovic2019geometric, xiao2021adiabatically, ren2021phonon}. This part is expressed as~\cite{trifunovic2019geometric} 
\begin{align} \label{theta0}
    M_z^t &=  \frac{-e}{T} \frac{1}{2\pi} \theta 
\end{align}
where $-e$ is the elementary electron charge, $T$ is the spin precession period, and the coefficient $\theta$ is a Chern-Simons form
{\begin{align}\label{theta1}
    \theta = \frac{1}{4\pi} \int_0^{2\pi} d\tau \int d^2k {\rm Tr} 
    \left[ \epsilon_{ijk} (A_i \partial_j A_k - \frac{2i}{3} A_i A_j A_k)
    \right] 
\end{align}
where $\epsilon_{ijk}$ is the Levi-Civita symbol with summation over repeated indices, ${\rm Tr}$ means trace over the occupied bands, $\tau = \omega t$ with $\omega=2\pi/T$, and $\bm{A}=(A_{k_x}, A_{k_y}, A_\tau)$.} 
$A_i$ is nonabelian Berry connection and its matrix element $A_i^{mn}(\bm{k})=\langle m,\bm{k} | i\partial_i |n, \bm{k}\rangle$ where $|m, \bm{k}\rangle$ and $|n, \bm{k}\rangle$ are electronic states of occupied bands at momentum $\bm{k}$ with band indices $m,n$. {Since the integral is gauge invariant only modulo $2\pi$, the pumped orbital magnetization can differ by an integer multiple of $e/T$ depending on the gauge choice. As shown in the Supplemental Materials~\cite{SM}, this uncertainty manifests in the boundary contributions. Specifically, depending on the crystalline orientation of the edge, the boundary contribution can vary by a quantized current resulting from the Thouless pumping.}


The gauge dependence of the integrand makes it difficult to evaluate $\theta$ directly. One efficient way to evaluate $\theta$ is to treat it as a function of an auxiliary parameter $\alpha$~\cite{xiao2009polarization, essin2009magnetoelectric}. 
Following the fundamental theorem of calculus, 
\begin{align}
    \theta (\alpha) - \theta (0) 
    =& \int_0^{\alpha} d \alpha' \frac{\partial \theta}{\partial \alpha'}.
\end{align}
When the Berry connections $\bm{A}$ can be made smooth and periodic in the Brillouin zone, the equation becomes~\cite{xiao2009polarization, essin2009magnetoelectric}
\begin{align} \label{theta2}
\theta (\alpha) - \theta (0) 
= \frac{1}{2\pi} \int_0^{\alpha} d\alpha' \int d\tau d^2k ~ {\rm Tr} \Omega_{k_xk_y\alpha'\tau}
\end{align}
and 
\begin{align} \label{secondChernForm}
\Omega_{k_xk_y\alpha'\tau}=\Omega_{k_x k_y}\Omega_{\alpha'\tau} + \Omega_{k_x \tau}\Omega_{k_y\alpha'} - \Omega_{k_x \alpha'}\Omega_{k_y\tau},
\end{align}
where the nonabelian Berry curvature $\Omega_{ij}=\partial_i A_j - \partial_j A_i - i[A_i, A_j]$. 
When the starting point $\theta(0)$ is chosen to be zero, one can obtain the accurate value of $\theta(\alpha)$ as Eq.~\eqref{theta2} is gauge invariant and path independent. 

Besides the pumped orbital magnetization, the topological magnetoelectric coupling constant is also expressed as a Chern-Simons form~\cite{qi2008topological, xiao2009polarization, essin2009magnetoelectric, mong2010antiferromagnetic, he2022topological}. This coupling is notably half-quantized, protected by either time-reversal or inversion symmetry, because the coupling represents the response of a pseudovector to a vector or vise versa, where the inversion or time-reversal operation only changes one of their signs. However, these symmetries cannot make the orbital magnetization quantized as the Chern-Simons form in Eq.~\eqref{theta1} represents the response of a pseudo-vector to the spin precession characterized by another pseudovector, $\bm{S}\times \dot{\bm{S}}$ as defined below. 

\textit{\textbf{Precession angle dependence.---}}{By defining the auxiliary parameter as spin precession angle $\alpha$ defined in Fig.~\ref{Fig4}(a), we can obtain the pumped orbital magnetization as a function of spin precession angle. 
For simplicity, we assume that $\bm{S}_A=-\bm{S}_B=\bm{S}$ during the spin precession as these two spin sublattices are locked by strong exchange interaction that is much stronger than the anisotropy. The spin orientations are denoted by a unit vector $\bm{S}=(\sin \alpha \cos \omega t, \sin \alpha \sin \omega t, \cos \alpha)$.}
Figure~\ref{Fig4}(b) plots the pumped orbital magnetization as a function of the precession angle following Eq.~\eqref{theta2}. 
When $\alpha=0$, the spin orientation is actually fixed and thus the pumped orbital magnetization vanishes. As $\alpha$ increases, the pumped orbital magnetization increases. 

\begin{figure}
\includegraphics[width=1\columnwidth]{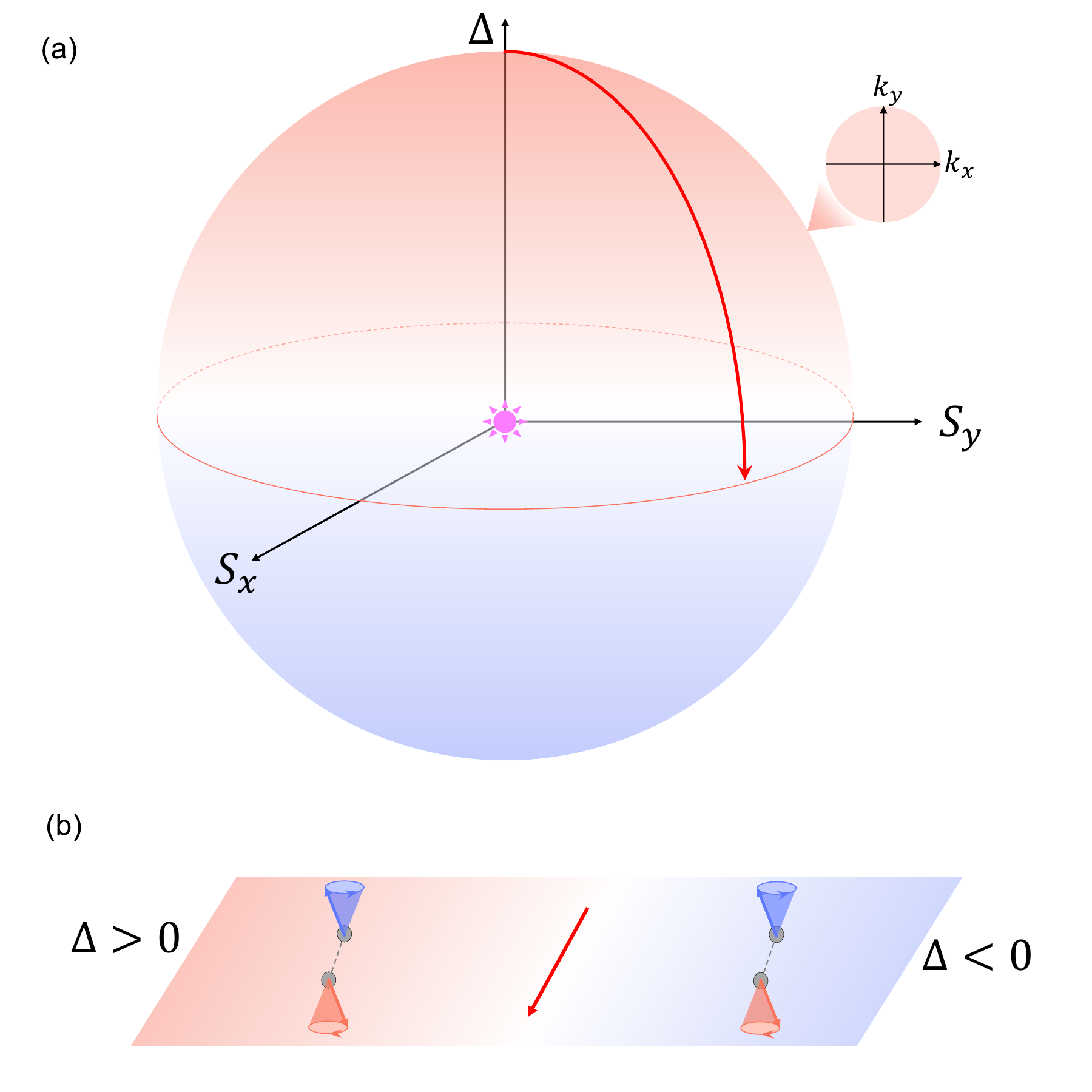}
\caption{(a) Schematic of a Yang's monopole in a five-dimensional parameter space spanned by $(k_x, k_y, \Delta, S_x, S_y)$. (b) A realization of Yang's monopole in real-space domain wall.}
\label{Fig5}
\end{figure}

When the angle is small, $S_{x,y}$ are much smaller than 1. Their influences on the energy band and the eigenstates are perturbative. {The Chern-Simons form in Eq.~\eqref{theta1} can thus be evaluated by series expansion with respect to $S_{x,y}$~\cite{ren2021phonon}. The orbital magnetization reads $M \simeq \chi (\bm{S}\times \dot{\bm{S}})_z$ with a constant 
\begin{align} \label{SecondChern}
    \chi = \frac{e}{2}\int \frac{d^2k}{(2\pi)^2} \Omega_{k_x k_y S_x S_y}
\end{align}
where the second Chern form in the integrand are evaluated at ${S_{x,y}=0}$. Therefore, the pumped orbital magnetization is proportional to $\alpha^2$ through $(\bm{S}\times \dot{\bm{S}})_z$. The magnetization depends on the spin precession chirality while is independent of the N\'eel vector. This dynamically induced magnetization corrects the Landau free energy in the presence of an external $\bm{B}$ field by $-\chi \bm{B}\cdot (\bm{S}\times \dot{\bm{S}})$ leading to a dynamical contribution to the magnetic susceptibility. The energy correction also manifests in the $\bm{B}$ field induced energy shift of magnons, i.e., the magnon Zeeman splitting, considering the magnetization induced by a single magnon.

The orbital magnetization induced by pumping processes is intricately linked to a range of magnetoelectric phenomena as suggested by Eq.~\eqref{SecondChern}. $\Omega_{k_x k_y S_x S_y}$ involves three types of Berry curvature, i.e., $\Omega_{k_i S_j}$, $\Omega_{k_x k_y}$, and $\Omega_{S_x S_y}$. $\Omega_{k_i S_j}$ is closely related to the spin-driven ferroelectricity. Specifically, the change of the polarization along $i$-th direction induced by the deviation in spin, $\delta S_j$, is proportional to $\delta S_j \cdot \int d^2k \Omega_{k_i S_j}$. As a result, a spatially varying spin structure leads to a nonzero charge density~\cite{brey1995skyrme, freimuth2013phase} while temporal precession of spins generates rotating polarization and thus orbital magnetization. 
Unlike $\Omega_{k_xk_y}$, which is closely related to various Hall effects, $\Omega_{S_xS_y}$ is associated with spin dynamics. In specific, $\Omega_{S_xS_y}$ 
characterizes the geometric phase of valence electrons that evolve adiabatically following spin precession. This phase leads to a geometric torque on the spins~\cite{michel2024bound}, enabling the control of spin dynamics by engineering valence electrons. The effects of such geometric torque are not yet fully understood.} 

As the spin precession cone angle increases, the orbital magnetization increases rapidly and can reach the order of $e/T$ as $\theta$ reaches $-\pi$. 
The large pumping has a topological origin. To show this, we consider the low-energy effective model near the K$'$ point, where massive Dirac dispersion appears described by the effective Hamiltonian $H_{K'}=\bm{q}\cdot \bm{\Gamma}$ where $\bm{q}=(v_Fk_x,v_Fk_y,-\Delta,\lambda S_x,\lambda S_y)$, $\Delta=3\sqrt{3}\gamma - \lambda S_z$, and the matrices $\bm{\Gamma}=(\sigma_x s_0, \sigma_y s_0, \sigma_z s_z, \sigma_z s_x, \sigma_z s_y)$. The energy bands are doubly degenerate with eigenenergy $E_\pm = \pm q$ and $q=|\bm{q}|$. When the spin $\bm{S}$ is along $z$ direction, the band gap is determined by $\Delta$, as shown in Fig.~\ref{Fig4}(f), and a single Dirac cone appears at $\Delta=0$ by reducing $S_z$ as shown in Fig.~\ref{Fig4}(e). The dispersion is gapped again if there is in-plane spin components as shown in Fig.~\ref{Fig4}(d). The second Chern form in Eq.~\eqref{theta2} resembles the nonabelian gauge field created by a Yang's monopole in a five-dimensional space around the gap closing point as illustrated in Fig.~\ref{Fig5}(a). Integrating the second Chern form over a closed surface containing the gap closing point leads to an integer second Chern number $C_2$ with $\theta=2\pi C_2$ whereas the integration over part of a closed surface, see the red arrow in Fig.~\ref{Fig5}(a), leads to a fraction of $2\pi C_2$.

{\textit{\textbf{Inhomogeneity enabled current pumping.---}}}
{The pumping of orbital magnetization facilitates the spin texture-enabled generation of an adiabatic charge current. Specifically, when orbital magnetization is spatially inhomogeneous, it generates a dissipationless charge current density, described by $\bm{j} = \nabla \times M(\bm{r}) \hat{z}$. In antiferromagnets, these inhomogeneities are often a result of spatial variations in spin textures. As the magnetic order varies across different regions, the pumped orbital magnetization also varies, thus enabling the controlled generation of a charge current. 
Such a current can even become quantized as $-\frac{e}{T}C_2$ by spatially engineering the electronic band topology. In specific, the electronic band gap $\Delta$ is determined by the competition between the spin-orbit coupling $\gamma$ and effective exchange field $\lambda$ from the magnetic order. 
By engineering a continuous change of $\Delta$ from positive to negative, one can map the parameter space shown in Fig.~\ref{Fig5}(a) to a domain wall structure in Fig.~\ref{Fig5}(b). By further introducing a spin precession enclosing a Yang's monopole inside the parameter space, a quantized magnetization current along the domain wall appears that is experimentally measurable.}

\begin{figure}
\includegraphics[width=1.\columnwidth]{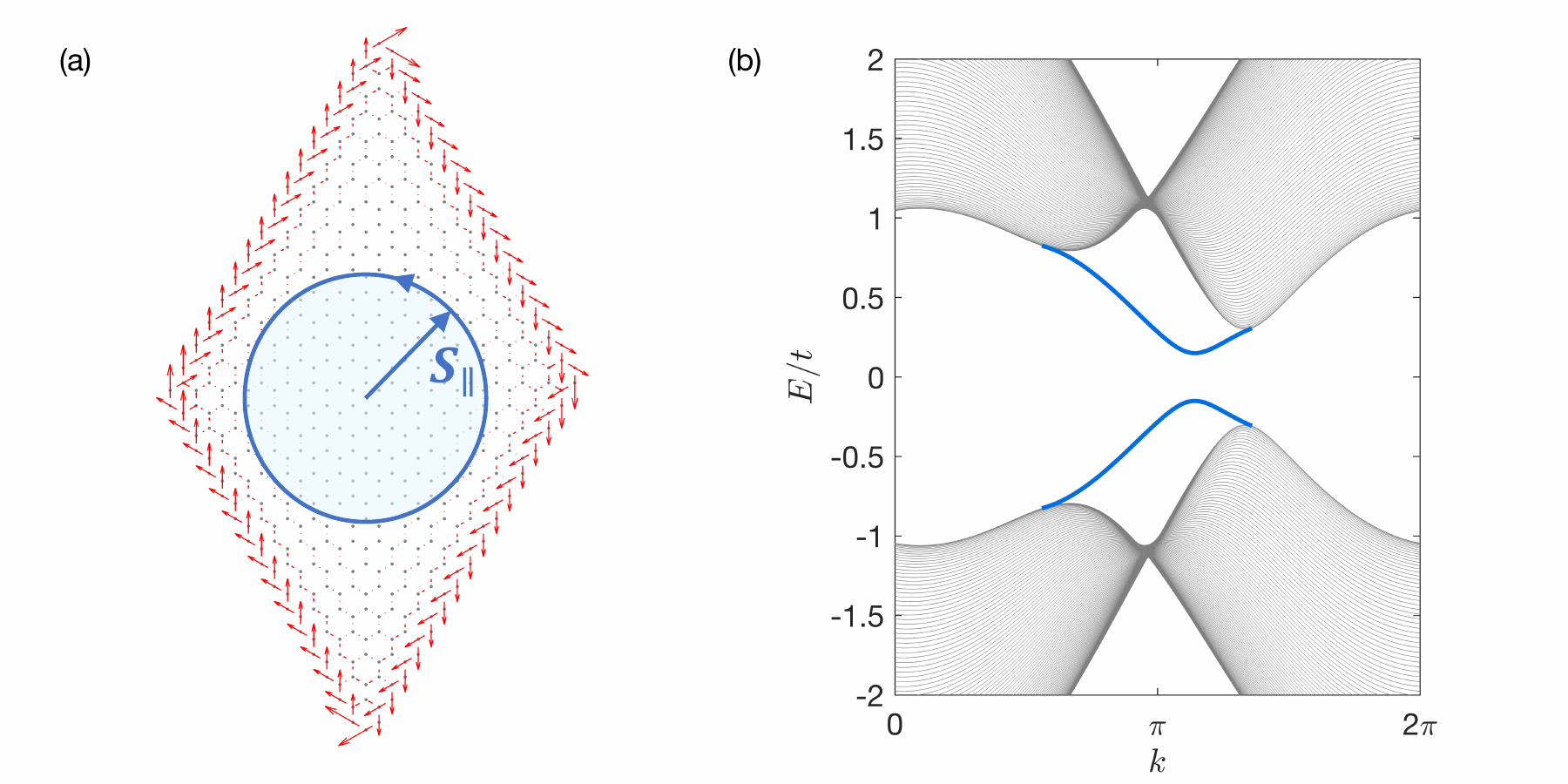}
\caption{(a) Schematic of the orbital magnetization pumping. The spin $\bm{S}$ precesses in the $x$-$y$ plane that induces a loop current along the boundary, thus an orbital magnetization. Red arrows are spatial maps of current in the lattice model. (b) The energy spectrum of a zigzag terminated nanoribbon of the Kane-Mele-Hubbard model with N\'eel vector tilted away from the $z$ direction. Blue bands denote helical edge states.}
\label{FigAFM1}
\end{figure}

\textbf{\textit{Boundary contribution.---}} {Besides the bulk, the boundary can also have significant contribution to the pumped orbital magnetization. When the N\'eel vector points out-of-plane, the antiferromagnetic topological insulator can host a quantum spin Hall effect with gapless edge states when $|3\sqrt{3}\lambda|>|\lambda S_z|$. The helical edge states described by the effective Hamiltonian $H_{\rm edge}=v k s_z$ where $k$ is the momentum along the boundary, $v$ is the velocity, and $\bm{s}$ are the spin Pauli matrices. By tilting the N\'eel vector, the edge states become gapped by a mass term correction $H_{\rm edge}'= m \bm{S}_{\parallel}\cdot \bm{s}$ where $m$ represents the effective exchange coupling strength. 
When the spin starts to precess with a frequency of $\omega=2\pi/T$ and $\bm{S}_{\parallel}={S}_{\parallel}(\cos \omega t, \sin \omega t)$, a quantized Thouless pumping along the boundary is realized if the adiabatic approximation is valid with $\hbar \omega \ll |2m {S}_{\parallel}|$. The current along the edge is  $J_{\rm edge}=C_1 e/T$ with $C_1$ being the first Chern number in the parameter space of $k$-$t$~\cite{thouless1983quantization, qi2008fractional} as illustrated by red arrows in Fig.~\ref{Fig1}(a). This current contributes an out-of-plane orbital magnetization $M_z=J_{\rm edge}$, which is even larger than the bulk contribution.}

Nevertheless, the boundary contribution is notably sensitive to the boundary conditions. {Firstly, the edge state properties can depend on the crystalline orientation, which remain gapless along certain orientations even in the presence of spin precession while become gapped along others. One example in the Supplemental Materials demonstrates a quantized Thouless pumping along the zigzag boundary, which is absent along the armchair boundary. Secondly, in antiferromagnetic topological insulators, the edge states are not protected by symmetry. Disorders on the boundary can destroy the quantized Thouless pumping.

The sensitivity of boundary contribution suggests that the pumped orbital magnetization can modulated by integer multiples of $e/T$ by engineering the boundary without altering the bulk. This uncertainty of orbital magnetization is a physical manifestation of the gauge dependence of Eq.~\eqref{theta1} where $\theta$ can vary by $2\pi$ with different gauge choices. Such uncertainty mirrors that of bulk polarization~\cite{king1993theory, vanderbilt1993electric}, which is also defined by pumping.}

\textit{\textbf{Summary.---}} 
We have studied the adiabatic pumping of orbital magnetization drive by spin precession in magnetic topological systems. {This pumping mechanism enables the rectification of spin precession, resulting in a static orbital magnetization or a dissipationless charge current in the presence of inhomogeneities such as spin textures or electronic topological domains. The orbital magnetization contributes to the dynamical magnetic susceptibility as well as the magnon Zeeman effect when the precession cone angle is small. For large cone angles, the orbital magnetization is related to the global properties of the electronic geometric phases in the parameter spaces spanned by momentum $\bm{k}$ and spin $\bm{S}$. In addition, a nontrivial quantized boundary contribution may emerge, illustrating the quantum uncertainty associated with the Chern-Simons form in bulk contribution.

The Kane-Mele Hubbard model serves as a prototype model hosting antiferromagnetic topological insulator, which has attracted many attentions~\cite{rachel2018interacting} and can be realized in moir\'e systems, like twisted MoTe$_2$~\cite{liu2024gate}. To experimentally measure the magnetization, the temperature should be lower than the N\'eel temperature and the electronic band gap. The order of magnitude of the pumped magnetization and adiabatic current can vary widely but can reach the order of $e/T$ with a large precession angle. The pumped magnetic moment per unit cell is proportional to the precession period and the unit-cell size, potentially reaching the order of a Bohr magneton in a moir\'e supercell about 50 nm$^2$ in area with terahertz frequency precession. The orbital-magnetization pumping is universal in both antiferro- and ferromagnetic systems~\cite{SM}. We expect similar phenomena in MnBi$_2$Se$_4$ systems with easy-plane anisotropy~\cite{chen2024antiferromagnetic}. Enhancements in this pumping effect are facilitated by strong spin-orbit coupling, narrow energy gap, nontrivial band topology, fast spin dynamics, and large cone angle of precession.}

\begin{acknowledgments}
\textit{Acknowledgments.---} We thank the helpful discussion with Qian Niu, Jiang Xiao, and Ran Cheng. The analytical study is supported by DOE Award No. DE-SC0012509, and the numerical simulation is supported by the Center on Programmable Quantum Materials, an Energy Frontier Research Center funded by DOE BES under award DE-SC0019443. Y.R. acknowledges startup funds provided by the College of Arts and Sciences and the Department of Physics and Astronomy of the University of Delaware.
\end{acknowledgments}

\end{document}